\documentclass[aps,pra,preprint,showpacs]{revtex4-1}

\usepackage{amsmath}
\usepackage{amssymb}
\usepackage{graphicx}
\usepackage{color}

\begin{document}

\title{Magneto-optical trasnport properties of monolayer WSe$_{2}$}
\author{M. Tahir$^{\star}$ and P. Vasilopoulos$^{\dag}$}
\affiliation{$^{1}$Department of Physics, Concordia University, Montreal, Quebec, Canada H3G 1M8}

\begin{abstract}
The recent experimental realization of a high quality WSe$%
_{2} $ leads to the possibility of magneto-optical measurements and
the manipulation of the spin and valley degrees of freedom. We study the influence 
of the very strong spin-orbit coupling 
and  of the anisotropic lifting of the  valley pseudospin
degeneracy  on  its magnetotransport properties.  
The energy spectrum of WSe$_{2} $ is derived and discussed
in the presence of a perpendicular magnetic field $B$. 
Correspondingly we evaluate  the magneto-optical Hall conductivity and  the optical longitudinal
conductivity as functions of the frequency, magnetic field, and Fermi
energy.  They are strongly influenced by the field $B$  and the strong spin splitting.  
The former exhibits valley polarization and the latter beatings of oscillations.
The  magneto-optical responses can be tuned in two
different regimes:  the mictrowave-to-terahertz regime and the visible-frequency one.
The absorption peaks involving the $n=0$ LL appear in between these two regimes and
show a magnetic control of the spin and valley splittings.
We also evaluate the power absorption spectrum.
\end{abstract}

\maketitle

\section{Introduction}

Graphene possesses
extraordinary properties but its application to device fabrication is limited
by its zero band gap which makes graphene transistors suffer from a low on-off
current ratio  \cite{LY}. This has led to intensive
investigation of alternative materials with a finite band gap including
silicene \cite{JT}, germanene \cite{ML}, and the group VI transition-metal
dichalcogenides  MX$_{2}$, M=Mo,W; X=S,Se, \cite{QK,BA,DG,HW,XF}. This MX$_{2}$ 
family is an 
intriguing class of semiconductors when thinned down to
monolayers. The valence and conduction band extrema are located at both $K$
and $K^{\prime }$ points at the corners of the hexagonal Brillouin zone. The 
$K$ and $K^{\prime }$ points are related to each other by time-reversal 
symmetry and give rise to the valley degree of freedom of the
band-edge electrons and holes \cite{AL,JM,HS,ER}. It has been demonstrated that
a monolayer of MoS$_{2}$ has reasonable inplane carrier mobility, high thermal
stability, and good compatibility with standard semiconductor manufacturing 
\cite{BA}. These  properties render monolayer MoS$_{2}$ a
promising candidate for a wide range of applications, including
photoluminescence at visible wavelengths \cite{AL,QKA}, photodetectors with
high responsivity \cite{HS}, and field-effect transistors \cite{BA,HA,YK}.

Compared to MoS$_{2}$ the material WSe$_{2}$ has a much stronger spin-orbit-coupling (SOC): in the
valence band it is 2$\lambda _{v}^{\prime }=450$ meV and in the conduction
band 2$\lambda _{c}^{\prime }=30$ meV. This and its high quality  provide an excellent system for
spin and valley control \cite{GZ,AM}.
A high-mobility WSe$_{2}$ transistor has been demonstrated  at
room temperature \cite{HAS}. Although WSe$_{2}$ is a
direct-bandgap semiconductor (2$\Delta =1.7$ eV), the lifting of the valley
degeneracy allows for optical manipulation of the electron valley index. This
has been realized by applying a  magnetic field normal  
 to the two-dimensional (2D) layer, see Refs. \onlinecite{GZ,AM} which clearly 
demonstrate  the lifting of the valley degeneracy  in WSe$_{2}$. This is achieved by monitoring the energy
splitting between the two circularly polarized luminescence components, $%
\sigma ^{+}$ and $\sigma ^{-}$, associated with optical recombination in the two valleys. 

References \onlinecite{GZ,AM} studied  optical transitions in a monolayer of WSe$_{2}$ and related
compounds in magnetic fields. Direct optical transitions in a WSe$%
_{2}$ monolayer occur at the edge of the Brillouin zone, which mainly consists of
strongly localized d-orbitals of the transition metal. This is in contrast
with GaAs and other conventional semiconductors used in optoelectronics in
which the direct optical bandgap is situated at the centre of the Brillouin
zone. In a WSe$_{2}$ monolayer  there are several possible contributions to the
Zeeman splitting as the emission of circularly polarized light originates
from states with contrasting valley index, spin and orbital magnetic moment.
As the valleys can  be selectively addressed, these experiments allow the
different contributions to the Zeeman splitting to be determined.  
A magneto-optical investigation in high-quality samples of WSe$_{2}$ appeared in  Ref. \onlinecite{AP}.

Optical transport properties have been evaluated for graphene and  a
good agreement exists between theory and experiment \cite{VS}. Magneto-optical
properties of topological insulators (TIs) \cite{WA} and other single-layer
materials,  such as MoS$_{2}$ \cite{ZJ} and silicene \cite{LC}, have also been
investigated.   Several  properties of WSe$_{2}$ have been studied at  zero
magnetic field \cite{GZ,AM}. In a finite magnetic field though Landau levels (LLs) are formed  and transitions between them generate specific absorption lines in the magneto-optical conductivity. We are aware though only
of the experimental work \cite{AP}  but of no theoretical one on   
the magnetotransport  properties of WSe$_{2}$.  Accordingly,  studying these properties is 
 timely and expected to increase our understanding of this material. Further, 
WSe$_{2}$ is expected to show strong spin- and valley-controlled properties \cite{GZ,AM} in contrast to graphene. As will be shown, an important difference with it  and other
2D systems, in which the magneto-optical response occurs in the terahertz
(THz) regime, is that in WSe$_{2}$ it can be tuned to the microwave-to-THz and
visible-frequency ranges. This is similar to phosphorene's response \cite{MP}.

In this  work we study the ac magnetotransport properties of a WSe$_{2}$ monolayer  in a  
perpendicular magnetic field $B$. Using the spectrum of this material and general, Kubo-type formulas, expressed
explicitly in terms of single-particle eigenstates and eigenvalues,  we evaluate  the 
Hall and longitudinal conductivities as well as the  absorption spectrum. In Sec. II we present the basics of the model, in Sec. III the conductivity expressions  and limited theoretical calculations, and in Sec. III numerical results. 
We summarize in Sec. V. 
\section{ Model}  

We consider a monolayer of WSe$_{2}$ in the $(x,y)$ plane in the presence of
intrinsic SOC, spin and valley Zeeman fields, and a normal magnetic field $B$. Extending the 2D,
Dirac-type Hamiltonian of WSe$_{2}$  for $B=0$ \cite{AH}, we have
\begin{equation}
H^{s\eta }=v_{F}(\eta \sigma _{x}\Pi _{x}+\sigma _{y}\Pi _{y})+\Delta \sigma
_{z}+\eta s(\lambda _{c}\sigma _{+}+\lambda _{v}\sigma _{-})+sM_{z}-\eta
M_{v}  \label{1}
\end{equation}
Here $\eta =\pm 1$ for valleys $K$ and $K^{\prime }$, $\Delta $ is the mass
term that breaks the inversion symmetry, $\lambda _{c}=\lambda _{c^{\prime
}}/2,\lambda _{v}=\lambda _{v}^{\prime }/2$, ($\sigma _{x}$, $\sigma _{y}$, $%
\sigma _{z}$) are the Pauli matrices for the valence and conduction bands, $%
\sigma _{\pm }=\sigma _{0}\pm \sigma _{z}$, 
$v_{F}$ ($5\times 10^{5}$
m/s) is  the Fermi velocity, and $s_{z}=+1 (-1)$ is the up (down) spin. 
Further, $M_{z}=g^{\prime }\mu
_{B}B/2 $ is the Zeeman exchange field induced by ferromagnetic order, $%
g^{\prime }$ the Land\'{e} $g$-factor ($g^{\prime }=g_{e}^{\prime
}+g_{s}^{\prime }$), and $\mu _{B}$\ the Bohr magneton \cite{GZ,AM}. Also, $%
g_{e}^{\prime }=2$ is the free-electron $g$ factor and $g_{s}^{\prime }=0.21$
is the out-of-plane factor due to the strong SOC. The last
term, $M_{v}=g_{v}^{\prime }\mu _{B}B/2$, breaks the valley symmetry of the
levels and $g_{v}^{\prime }$ = 4 \cite{GZ,AM}. Further,  $\mathbf{\Pi =p}+e%
\mathbf{A}$ is the 2D canonical momentum with vector potential $%
\mathbf{A}$. Using the Landau gauge with $\mathbf{A}= (0, Bx,
0)$ and diagonalizing the Hamiltonian  (1) gives   the eigenvalues
\begin{equation}
E_{n}^{s\eta ,\gamma }=s\eta (\lambda _{c}+\lambda _{v})+sM_{z}-\eta
M_{v}+\gamma E_{n}^{s\eta },  \label{2}
\end{equation}
where $E_{n}^{s\eta }=[n\hslash ^{2}\omega _{c}^{2}+\Delta _{s\eta }^{2}%
]^{1/2}$, $\omega _{c}=
(2eB/\hslash)^{1/2}$ is the cyclotron frequency, $\gamma
=\pm 1$ represents electron and hole states, respectively, and $\Delta _{s\eta
}=\Delta +s\eta (\lambda _{c}-\lambda _{v})$. 
The eigenvalues (2) become  simpler upon noticing the inequality $\hslash \omega _{c}\ll \Delta
_{s\eta }$. Expanding the square root in $E_{n}^{s\eta }$ gives
\begin{equation}
E_{n}^{s\eta ,\gamma }\approx s\eta (\lambda _{c}+\lambda _{v})+sM_{z}-\eta
M_{v}+\gamma \Delta _{s\eta }+\gamma n\,\frac{\hslash ^{2}\omega _{c}^{2}}{%
2\Delta _{s\eta }}.  \label{3}
\end{equation}
This is a usual, linear in $n$ and $B$ LL spectrum. Using $\Delta _{s\eta
}\gg \eta s\lambda $ the last term is equal $\gamma n(\hslash ^{2}\omega
_{c}^{2}/2\Delta )(1+\eta \lambda )$. This gives a spin slitting $%
E(s=1)-E(s=-1)=2M_{z}+n\eta \lambda (\hslash ^{2}\omega _{c}^{2}/\Delta )$
in the conduction band and $2\eta \lambda -n\eta \lambda (\hslash ^{2}\omega
_{c}^{2}/\Delta )$ in the valence band. The term $n(\hslash ^{2}\omega
_{c}^{2}/2\Delta )\propto nB$ is about twice as big as $M_{z}$ and much
smaller than $\lambda $. It's important in the conduction band but
negligible in the valence band in which $\lambda \approx 450$ meV.
For very weak fields $B$ the linear dispersion, due to the huge band gap, has been discussed in Refs.\cite{QK,BA,DG,HW,XF,AH}.

The  eigenfunctions corresponding to Eq. (2) are obtained as
\begin{equation}
\Psi _{n}^{s\eta ,\gamma }=\frac{e^{ik_{y}y}}{\sqrt{L_{y}}}\left (
\begin{array}{c}
\eta C_{n}^{s\eta ,\gamma }\phi _{n}
\ \\
D_{n}^{s\eta ,\gamma }\phi _{n-1}
\end{array}\right),
\label{4}
\end{equation}
where $C_{n}^{s\eta ,\gamma }=\sqrt{n}\hslash \omega _{c}/
[n\hslash
^{2}\omega _{c}^{2}+(\Delta _{s\eta }-\gamma E_{n}^{s\eta })^{2}]^{1/2}$ and $%
D_{n}^{s\eta ,\gamma }=(-\Delta _{s\eta }+\gamma E_{n}^{s\eta })/
[n\hslash ^{2}\omega _{c}^{2}+(\Delta _{s\eta }-\gamma E_{n}^{s\eta })^{2}]^{1/2}$; $\phi_n(x)$ are 
harmonic oscillator functions. %
The eigenvalues  for $n=0$ 
are
\begin{equation}
E_{0}^{s,+}=\Delta +2s\lambda _{c}+sM_{z}-M_{v},\quad E_{0}^{s,-}=-\Delta
-2s\lambda _{v}+sM_{z}+M_{v}  \label{5}
\end{equation}
and the corresponding eigenfunctions 
\begin{equation}
\Psi _{0}^{s,+}=\frac{e^{ik_{y}y}}{\sqrt{L_{y}}}\left( 
\begin{array}{c}
\phi _{0} \\ 
0%
\end{array}%
\right),\quad \Psi _{0}^{s,-}=\frac{e^{ik_{y}y}}{\sqrt{L_{y}}}\left( 
\begin{array}{c}
0 \\ 
\phi _{0}%
\end{array}%
\right).  \label{6}
\end{equation}
%
To better appreciate the spectrum (2)  
one can contrast it
with that for $B=0$ given by 
\begin{equation}
E_{p}^{s,\eta }=s\eta (\lambda _{c}+\lambda _{v})+sM_{z}-\eta M_{v}+\gamma 
\left[ v^{2}\hslash ^{2}k^{2}+\Delta _{s\eta }^{2}\right] ^{1/2}.  \label{7}
\end{equation}
Here $\gamma =1(-1)$ denotes the conduction (valence) band, $s=1(-1)$ is for
spin up (down), and $\eta =1(-1)$ for the $K$ ( $K^{\prime }$) valley.
Further, $k$ is the 2D wave vector. The spectrum (7) is shown in Fig. 1
versus $ka$ where $a=0.331$ nm is the lattice constant.

\begin{figure}[t]
\includegraphics[width=0.4\columnwidth,height=0.28\columnwidth
]{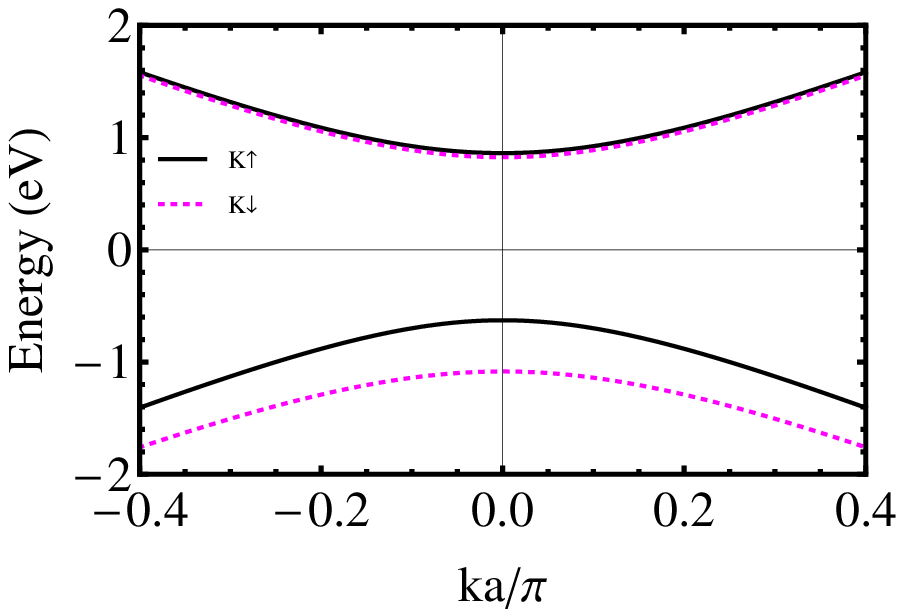}
\hspace*{0.6cm}
\includegraphics[width=0.4\columnwidth,height=0.28\columnwidth
]{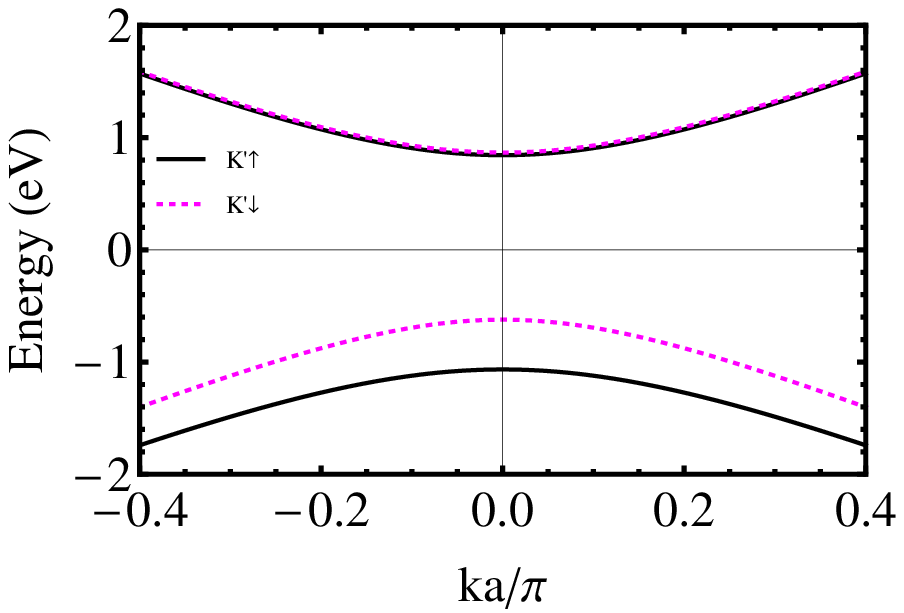}
\vspace*{-0.4cm}
\caption{(Coulour online) Band structure of WSe$_{2}$ in the  absence of a
magnetic field $B$.  The left panel is for the $K$ valley and the right one for the $K'$ valley.}
\end{figure}

We present the eigenvalues given by Eq. (2), as functions of the 
field $B$, in Fig. 2. The top and panel is for 
the conduction band and the bottom ones
for the valence band with finite spin $M_{z}$ and valley $M_{v}$
Zeeman fields. We find the following: (i) in contrast to the $\sqrt{B}$
dependence in graphene or silicene, the  LLs  grow linearly
with $B$. This is obvious from Eq. (3) which holds well because $\hslash
\omega _{c}\ll \Delta _{\eta s}$. (ii) For $M_{z}$ = $M_{v}=0$ the spin
splitting in the conduction band is enhanced due to last term in Eq. (3). It is
approximately an order of magnitude  larger than 
$M_{z}$ term and depends
linearly on the LL index $n$ and field $B$: for $n=5$ it is 33.3 meV at $B$ =
30 T. (iii) As Fig. 2 shows, the energies of the spin-up (down) LLs at the $%
K $ valley are different than those of the spin-down (up) at the $%
K^{^{\prime }}$ valley. This is in contrast
to MoS$_{2}$ where the spin splitting is negligible in the conduction band 
\cite{DG}. On the other hand, in the valence band the spin splitting is 450
meV and is the same as that for $B=0$. (vi) The $n=0$ LL is valley
degenerate for $M_{z}=$ $M_{v}=0$ in both the conduction and valence bands.
\begin{figure}[t]
\includegraphics[width=0.6\textwidth,clip]{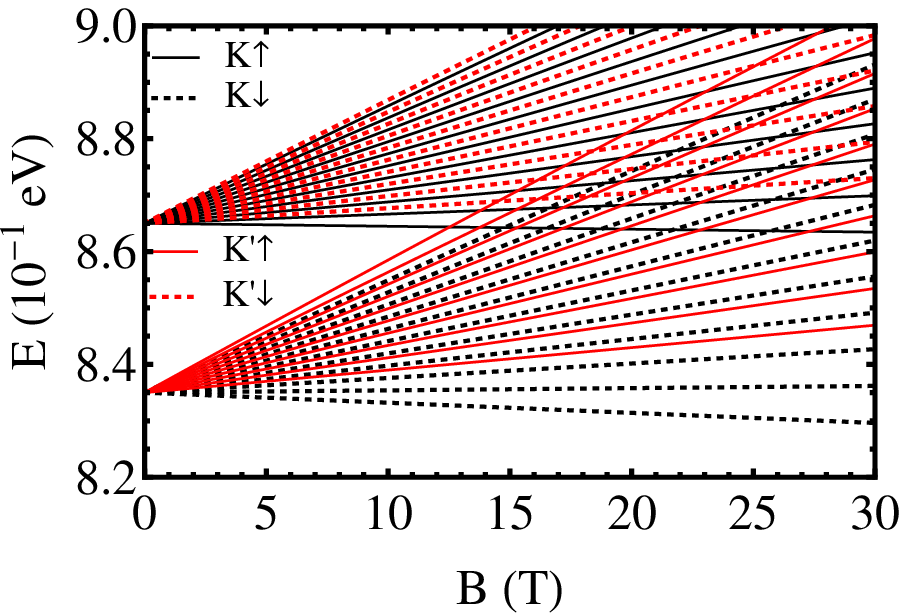}
\ \\%
\vspace{0.5cm}
\includegraphics[width=0.38\columnwidth,height=0.28\columnwidth]{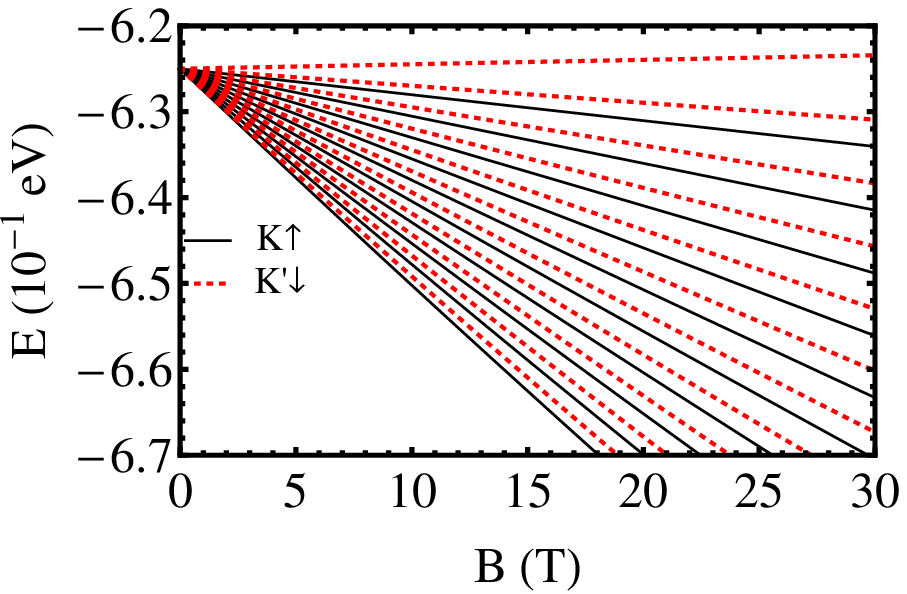}
\hspace{0.8cm}
\includegraphics[width=0.38\columnwidth,height=0.28\columnwidth]{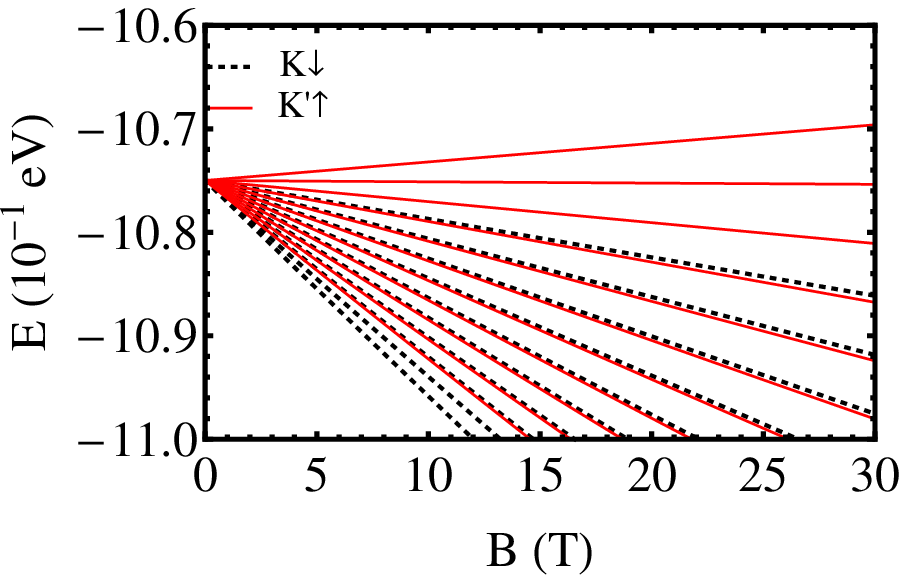}
\vspace*{-0.4cm}
\caption{(Coulour online) Band structure of MoS$_{2}$ versus 
magnetic field $B$ including spin and valley Zeeman fields. The top panel
is for the conduction band and the bottom ones for the valence band.}
\end{figure}

The density of states (DOS) is given by
\begin{equation}
D(E)=\frac{1}{S_{0}}\sum_{n,\eta ,s,k_{y}}\delta (E-E_{n}^{s\eta ,\gamma }),
\label{8}
\end{equation}
where $S_{0}=L_{x}L_{y}$ is area of the system. The sum over $k_{y}$ 
can be evaluated using the prescription ($k_{0}=L_{x}/2l^{2}$) $%
\sum_{k_{y}}\rightarrow (L_{y}/2\pi)g_{s}g_{v}%
\int_{-k_{0}}^{k_{0}}dk_{y}=(S_{0}/D_{0})g_{s}g_{v},$ $D_{0}=2\pi
l^{2}$; $g_{s}$ and $g_{v}$ are the spin and valley degeneracy factors. We
use $g_{s}$ = $g_{v}=1$ in the present work due to the lifting of the spin
and valley degeneracies. The Fermi level  $E_F$ is obtained from the electron
concentration $n_{c}$  given by 
\begin{equation}
n_{c}=\int_{-\infty }^{\infty }D(E)f(E)dE=(g_{s}/D_{0})\sum_{n,\eta
,s}f(E_{n}^{s\eta ,\gamma }),  \label{9}
\end{equation}
where $f(E_{n}^{s\eta
,\gamma })=(1+\exp [\beta (E_{n}^{s\eta ,\gamma }-E_F)])^{-1}$  is the Fermi-Dirac  
function with $\beta=1/k_{B}T$.
\begin{figure}[t]
\includegraphics[width=0.5\columnwidth,height=0.4\columnwidth]{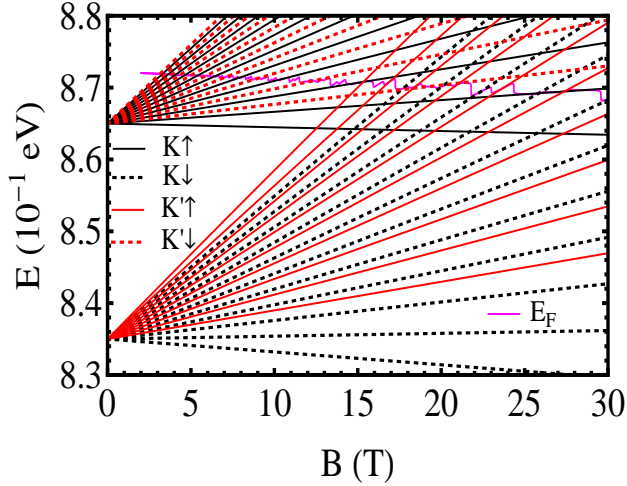}
\vspace*{-0.4cm}
\caption{(Coulour online) Fermi level of WSe$_{2}$ as a function of the  
magnetic field $B$ for $T$ = 1 K.}
\end{figure}

The magenta solid curve in Fig. 3 shows $E_F$, obtained
numerically from Eq. (9) for realistic values of $n_{c}=1\times 10^{17}$ m$%
^{-2}$, as a function of $B$; the LLs shown are the same as those in Fig. 2,
i.e., spin and valley dependent, since the magnetic field lifts the spin and
valley degeneracies of the $n\geq 1$ LLs . The additional intra-LL small
jumps result from the lifting of these 
degeneracies; the solid
and dashed curves ($n\geq 1$) are, respectively, for spins up and spins down
in the $K$ valley. For the $K^{\prime }$ valley the spins are
reversed, e.g., for $n\geq 1$, the spin-up electrons in the $K$ valley have
the same energy as the spin-down ones in the $K^{\prime }$ valley. For $%
n\geq 1$ the four-fold degeneracy, due to spin and valley, of all LLs is
lifted while the $n=0$ LL in the conduction band for $K$ valley and in the valence band for $K^{\prime }$ valley. The
results for $E_F$ in Fig. 3, with $M_{z}\neq  0$ and $M%
_{v} \neq  0$, correspond to the four-fold nondegenerate LLs. 
\begin{figure}[ht]
\hspace*{-0.5cm}
\includegraphics[width=0.5\columnwidth,height=0.3\columnwidth]{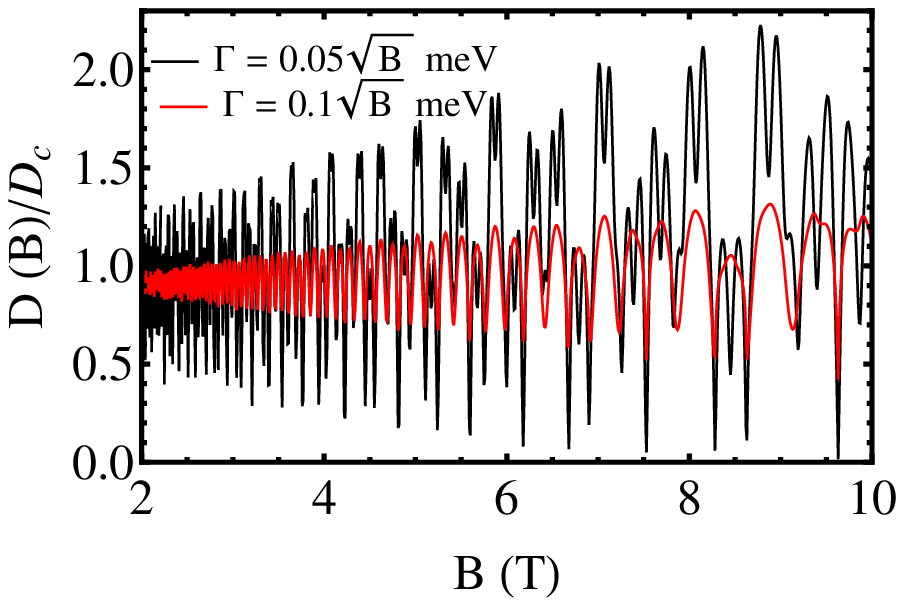}
\hspace*{0.7cm}
\includegraphics[width=0.4\columnwidth,height=0.3\columnwidth]{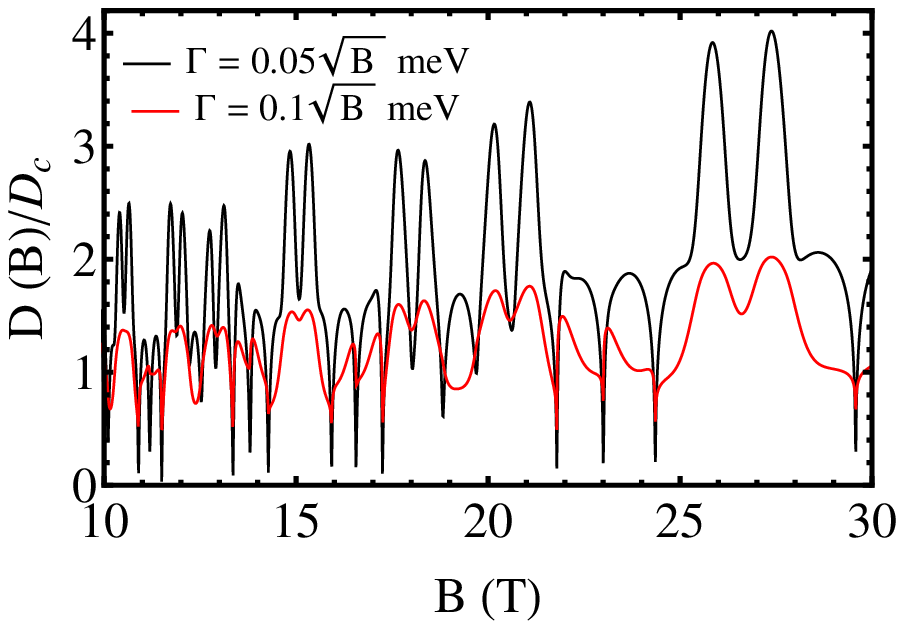}
\vspace*{-0.4cm}
\caption{(Coulour online) Dimensionless density of states as a function of the  
field $B$ for a LL width $\Gamma =0.05\sqrt{B}$\ meV (black curve) and $\Gamma =0.1\sqrt{B}$\ meV (red curve).
The two panels  differ only in the  range of the $B$ field 
($x$ axis).}
\end{figure}

Assuming a Gaussian broadening of the LLs, for zero temperature, the DOS per
unit area given in Eq. (8) is written as $D(E)=(g_{s}g_{v}/(D_{0}\Gamma 
\sqrt{2\pi }))\sum_{n,s}\exp \left[ -(E-E_{n}^{s\eta ,\gamma })^{2}/2\Gamma
^{2}\right]$, where $\Gamma $ is the width of the Gaussian distribution \cite%
{YT}. The DOS is shown in Fig. 4 as a function of the magnetic
field, $\Gamma =0.05\sqrt{B}$\ meV (solid black), $\Gamma =0.1\sqrt{B}$\ meV
(dotted red), and realistic values of $n_{c}=1\times 10^{17}$ m$^{-2}$ \cite%
{AP}. We obtained the SdH oscillations as a function of magnetic field in
the conduction band with equally spaced LLs. In weak magnetic fields $B$
the level broadening effect is significant due to the small LL separation 
whereas in strong fields it may become weaker due to  $\Gamma$'s 
dependence on the  field as $\sqrt{B%
}$ and the strong LL separation. 

As in the case of a 2DEG \cite{Wang}, the beating is due to 
the closeness of the frequencies of the spin-up and
spin-down states that result from the splitting of the LLs due to the SOC.
The beating shows up at low fields and the splitting of the oscillations
becomes more pronounced at high fields. The beating persists in the
conduction band for magnetic fields up to about 10 Tesla. Above this value
it is quenched and the SdH oscillations are split. This behaviour is
explained by the closeness of the oscillation frequencies of the SOC-split
LLs. The magnetic field-enhanced splitting in the conduction band mixes the
spin-up and spin-down states of neighbouring LLs into two unequally spaced
energy branches. The beating appears when the subband broadening is of the
order of $\hslash \omega _{c}$. For high magnetic fields the SOC effects
weaken and the beating pattern is replaced by a splitting of the peaks,
which persist due to the SOC and Zeeman energies

\section{Linear-response conductivity expressions}

We consider a many-body system described by the Hamiltonian $H=H_{0}+H_{I}-%
\mathbf{R}\cdot \mathbf{F}(t)$, where $H_{0}$ is the unperturbed part, $%
H_{I} $ is a binary-type interaction (e.g., between electrons and impurities
or phonons), and $-\mathbf{R}\cdot \mathbf{F}(t)$ is the interaction of the
system with the external field $F(t)$ \cite{MK}. For conductivity problems
we have $\mathbf{F}(t)=e\mathbf{E}(t)$, where $\mathbf{E}(t)$ is the
electric field, $e$ the electron charge, $\mathbf{R}=\sum_{\mathbf{r}_{i}}$,
and $\mathbf{r}_{i}$ is the position operator of electron $i$. In the
representation in which $H_{0}$ is diagonal the many-body density operator $%
\rho =\rho ^{d}+\rho ^{nd}$ has a diagonal part $\rho^{d}$ and a
nondiagonal part $\rho ^{nd}$. For weak electric fields and weak scattering
potentials, for which the first Born approximation applies, the conductivity
tensor has a diagonal part $\sigma _{\mu \nu }^{d}$ and a nondiagonal part $%
\sigma _{\mu \nu }^{nd}$ part, $\sigma _{\mu \nu }=\sigma _{\mu \nu
}^{d}+\sigma _{\mu \nu }^{nd}$, $\mu ,\nu =x,y$.

In general we have two kinds of currents, diffusive and hopping, with  $\sigma _{\mu \nu }^d=\sigma _{\mu \nu
}^{dif} +\sigma _{\mu \nu }^{col}$, but usually
only one of them is present.  When a magnetic field is present
we have only hopping current since the diffusive part $\sigma _{\mu \nu}^{dif} $
vanishes identically due to  vanishing velocity matrix elements  as is evident, 
for quasi-elastic scattering, 
by its form \cite{MK}
\begin{equation}
\sigma _{\mu \nu }^{d}(\omega )=\frac{\beta e^{2}}{S_{0}}\sum_{\zeta
}f_{\zeta }(1-f_{\zeta })\frac{v_{\nu \zeta }\,v_{\mu \zeta }\,\tau _{\zeta }%
}{1+i\omega \tau _{\xi }},  \label{10}
\end{equation}
where $\tau _{\zeta }$ is the momentum relaxation time, $\omega $ the
frequency, and $v_{\mu \zeta }$ the diagonal matrix elements of the velocity
operator. Further, $f_{\zeta }=[1+\exp {\beta (E_{\zeta }-E_{F})}]^{-1}$ is
the Fermi-Dirac distribution function,  $\beta =1/k_{B}T$, $T$ the
temperature, $E_{F}$ the Fermi level,  and $S_0$ the area of the sample.
In our case $v_{\mu \zeta }=0$  and  the conductivity given by Eq.
(10) vanishes. As for the ac hopping conductivity $\sigma _{\mu \nu }^{col}$, it is  given by Eq.
(2.64) of Ref. \cite{MK}, in strong fields $B$ is much smaller than the 
contribution $\sigma _{\mu \nu }^{nd} $ given below, and is neglected.

Regarding the contribution $\sigma _{\mu \nu }^{nd}$ one can use the
identity $f_{\zeta }(1-f_{\zeta ^{\prime }})[1-\exp {\beta (E_{\zeta
}-E_{\zeta ^{\prime }})}]=f_{\zeta }-f_{\zeta ^{\prime }}$ and cast the
original form  \cite{MK} in the more familiar one
\begin{equation}
\sigma _{\mu \nu }^{nd}(\omega )=\frac{i\hbar e^{2}}{S_{0}}\sum_{\zeta \neq
\zeta ^{\prime }}\frac{(f_{\zeta }-f_{\zeta ^{\prime }})\,v_{\nu \zeta \zeta
^{\prime }}\,v_{\mu \zeta ^{\prime }\zeta }}{(E_{\zeta }-E_{\zeta ^{\prime
}})(E_{\zeta }-E_{\zeta ^{\prime }}+\hbar \omega -i\Gamma_\zeta )}\,,  \label{11}
\end{equation}
\noindent where the sum runs over all quantum numbers $\left\vert \zeta
\right\rangle \equiv \left\vert n,s,k_{y}\right\rangle $ and $\left\vert
\zeta ^{\prime }\right\rangle \equiv \left\vert n^{\prime },s^{\prime
},k_{y}^{\prime }\right\rangle $ with 
$\zeta \neq \zeta ^{\prime }$. The
infinitesimal quantity $\epsilon $ in the original form \cite{MK} has been
replaced by $\Gamma _{\zeta }$ to account for the broadening of the energy
levels. Here $v_{\nu \zeta \zeta ^{\prime }}$ and $v_{\mu \zeta \zeta
^{\prime }}$ are the offdiagonal matrix elements of the velocity operator.
Using Eqs. (1) and (3) for the K-valley gives
\begin{equation}
v_{x,n,n^{\prime }}=v\big[C_{n}^{s\eta ,\gamma }D_{n^{\prime }}^{s\eta ,\gamma
^{\prime }}\delta _{n,n^{\prime }-1}+D_{n}^{s\eta ,\gamma }C_{n^{\prime
}}^{s\eta ,\gamma ^{\prime }}\delta _{n-1,n^{\prime }}\big] \delta _{k_y,k_y^\prime} \label{12}
\end{equation}
\begin{equation}
v_{y,n^{\prime },n}=-i\eta v\big[C_{n^{\prime }}^{s\eta ,\gamma ^{\prime
}}D_{n}^{s\eta ,\gamma }\delta _{n^{^{\prime }},n-1}-D_{n^{\prime }}^{s\eta
,\gamma ^{\prime }}C_{n}^{s\eta ,\gamma }\delta _{n^{^{\prime }}-1,n}\big] \delta _{k_y,k_y^\prime}
\label{13}
\end{equation}
Similarly, by exchanging $n$ with $n^{\prime }$ only in the Kronecker deltas
one obtains the results for the $K^{\prime }$ valley. Since $%
\left\vert \zeta \right\rangle \equiv \left\vert n,s,k_{y}\right\rangle $,
there will be one summation over $k_{y}$\ which, with periodic boundary
conditions for $k_{y}$, gives the factor $S_{0}/2\pi l^{2}$. 
As usual, the matrix elements between the $n=0$ LL and the other LLs
are treated separately. Using Eqs. (1) and (3), we arrive at%
\begin{equation}
v_{x,n,n^{\prime }}^{0}=v[D_{n^{\prime }}^{s\eta ,\gamma ^{\prime }}\delta
_{0,n^{\prime }-1}+D_{n}^{s\eta ,\gamma }\delta _{0,n-1}]  \label{14}
\end{equation}%
\begin{equation}
v_{y,n^{\prime },n}^{0}=-iv[D_{n^{\prime }}^{s\eta ,\gamma ^{\prime }}\delta
_{0,n^{\prime }-1}-D_{n}^{s\eta ,\gamma }\delta _{0,n-1}]  \label{15}
\end{equation}%
Similarly, by exchanging $D\rightarrow C$ in delta function only, we can be
obtained for the $K^{\prime }$-valley.
Using Eq. (12)
into Eq. (11), we obtain the longitudinal component $\sigma _{xx}^{nd}$  as
\begin{equation}
\sigma _{xx}^{nd}(\omega )=i\sigma _{0}\sum_{s,\eta ,n,n^{\prime },\gamma
,\gamma ^{\prime }}\frac{(f_{n}^{s\eta ,\gamma }-f_{n^{\prime
}}^{s\eta ,\gamma ^{\prime }}) \big[ D_{n}^{s\eta ,\gamma
}C_{n^{\prime }}^{s\eta ,\gamma ^{\prime }}\delta _{n-1,n^{\prime
}}+C_{n}^{s\eta ,\gamma }D_{n^{\prime }}^{s\eta ,\gamma ^{\prime }}\delta
_{n+1,n^{\prime }}\big] }{(E_{n}^{s\eta ,\gamma }-E_{n^{\prime }}^{s\eta
,\gamma ^{\prime }})(E_{n}^{s\eta ,\gamma }-E_{n^{\prime }}^{s\eta ,\gamma
^{\prime }}+\hbar \omega -i\Gamma )}.  \label{16}
\end{equation}
The  matrix elements of the velocity operators are nonzero only for
$n^{\prime }=n\pm 1$. Summing over $n^{\prime }$ and
seting $\sigma _{0}=\hbar e^{2}v_{F}^{2}/(2\pi l^{2})$ we arrive at
\begin{equation} 
\sigma _{xx}^{nd}(\omega ) =i\sigma _{0}\sum_{s,\eta ,n,\gamma ,\gamma
^{\prime }}\left [\frac{\big( f_{n}^{s\eta ,\gamma }-f_{n-1}^{s\eta ,\gamma
^{\prime }}\big) \,D_{n}^{s\eta ,\gamma }C_{n-1}^{s\eta ,\gamma ^{\prime }}%
}{I_{n,n-1}^{\gamma ,\gamma ^{\prime }}(I_{n,n-1}^{\gamma ,\gamma ^{\prime
}}+\hbar \omega -i\Gamma )}  
 \,\,
+\,\,%
\frac{\big (f_{n}^{s\eta ,\gamma }-f_{n+1}^{s\eta ,\gamma ^{\prime }}\big)
\,C_{n}^{s\eta ,\gamma }D_{n+1}^{s\eta ,\gamma ^{\prime }}}{%
I_{n,n+1}^{\gamma ,\gamma ^{\prime }}(I_{n,n+1}^{\gamma ,\gamma ^{\prime
}}+\hbar \omega -i\Gamma )}\right ],   
\end{equation} 
where $I_{n,n\pm 1}^{\gamma ,\gamma ^{\prime }}=E_{n}^{s\eta ,\gamma
}-E_{n\pm1}^{s\eta ,\gamma ^{\prime }}$. After making the changes $%
n-1\rightarrow m\rightarrow n$ in the first sum, we combine the two sums and
obtain
\begin{equation} 
\sigma _{xx}^{nd}(\omega ) =i\sigma _{0}\sum_{s,\eta ,n,\gamma ,\gamma
^{\prime }}\left [\frac{\left( f_{n+1}^{s\eta ,\gamma }-f_{n}^{s\eta ,\gamma
^{\prime }}\right) \,D_{n+1}^{s\eta ,\gamma }C_{n}^{s\eta ,\gamma ^{\prime }}%
}{I_{n+1,n}^{\gamma ,\gamma ^{\prime }}(I_{n+1,n}^{\gamma ,\gamma ^{\prime
}}+\hbar \omega -i\Gamma )}\,\,  
+\,\, \frac{\big(
f_{n}^{s\eta ,\gamma }-f_{n+1}^{s\eta ,\gamma ^{\prime }}\big)
\,C_{n}^{s\eta ,\gamma }D_{n+1}^{s\eta ,\gamma ^{\prime }}}{%
I_{n,n+1}^{\gamma ,\gamma ^{\prime }}(I_{n,n+1}^{\gamma ,\gamma ^{\prime
}}+\hbar \omega -i\Gamma )} \right ].  
\end{equation}%
In the limit $\Gamma \rightarrow 0,\,\omega \rightarrow 0$ and $\gamma
=\gamma ^{\prime }$ Eq. (16) yields zero. Now, one needs to sum over all
possible combinations of the matrix elements and for convenience we write $%
\sum_{\gamma ,\gamma ^{\prime }}=\sum_{+,+}+\sum_{-,-}+\sum_{+,-}+\sum_{-,+}$%
. Here the subscript $+/-$ denotes the conduction/valence band. After
performing the summation over $\gamma ,\gamma ^{\prime }$, we obtain  the
real part of $\sigma _{xx}^{nd}$ as  
\begin{align}
\notag 
\Re \sigma _{xx}^{nd}& =- \sigma _{0}\sum_{\eta ,s,n}\left\{ \frac{(
f_{n}^{s\eta ,+}-f_{n+1}^{s\eta ,+}) \,\Gamma \,\left( D_{n+1}^{s\eta
,+}C_{n}^{s\eta ,+}\right) ^{2}}{I_{n,n+1}^{+,+}\big[(I_{n,n+1}^{+,+}+\hbar
\omega )^{2}+\Gamma ^{2}\big]}\right.\,\,  
 +\,\,\frac{( f_{n}^{s\eta ,-}-f_{n+1}^{s\eta ,-})\,\Gamma \,\left(
D_{n+1}^{s\eta ,-}C_{n}^{s\eta ,-}\right) ^{2}}{%
I_{n,n+1}^{-,-}\big[(-I_{n,n+1}^{-,-}+\hbar \omega )^{2}+\Gamma ^{2}\big]}\\  
&\hspace{1.5cm} +\frac{( f_{n+1}^{s\eta ,-}-f_{n}^{s\eta ,+}) \,\Gamma \left(
D_{n+1}^{s\eta ,-}C_{n}^{s\eta ,+}\right) ^{2}}{%
I_{n+1,n}^{-,+}\big[(I_{n+1,n}^{-,+}+\hbar \omega )^{2}+\,\Gamma ^{2}\big]}  
 \left.\,\,+\,\,\frac{(f_{n}^{s\eta ,-}-f_{n+1}^{s\eta ,+}) \Gamma
\left( D_{n+1}^{s\eta ,+}C_{n}^{s\eta ,-}\right) ^{2}}{%
I_{n,n+1}^{-,+}\big[(I_{n,n+1}^{-,+}+\hbar \omega )^{2}+\Gamma ^{2}\big]}\right\} 
\end{align}%
Similarly,  exchanging $C$ with $D$ in Eq. (17) gives the results for the $K^{\prime }$ valley. 
Following the same procedure as opted
for the $n\geq 1$, we obtained the real part of the optical longitudinal
conductivity for the zeroth LL as
\begin{align}
\Re \sigma _{xx}^{nd}& =-\sigma _{0}\sum_{\eta ,s}\left\{ \frac{\left[
f_{0}^{s\eta ,+}-f_{1}^{s\eta ,+}\right] \,\Gamma \,\left( D_{1}^{s\eta
,+}\right) ^{2}}{I_{0,1}^{+,+}(I_{0,1}^{+,+}+\hbar \omega )^{2}+\Gamma ^{2}}%
\right.  \notag  
 +\frac{\left[ f_{0}^{s\eta ,-}-f_{1}^{s\eta ,-}\right] \,\Gamma \,\left(
D_{1}^{s\eta ,-}\right) ^{2}}{I_{0,1}^{-,-}(-I_{0,1}^{-,-}+\hbar \omega
)^{2}+\Gamma ^{2}}  \notag \\
&\hspace*{1.5cm} +\frac{\left[ f_{1}^{s\eta ,-}-f_{0}^{s\eta ,+}\right] \,\Gamma \left(
D_{1}^{s\eta ,-}\right) ^{2}}{I_{1,0}^{-,+}(I_{1,0}^{-,+}+\hbar \omega
)^{2}+\,\Gamma ^{2}}   
 \left. +\frac{\left[ f_{0}^{s\eta ,-}-f_{1}^{s\eta ,+}\right] \Gamma
\left( D_{1}^{s\eta ,+}\right) ^{2}}{I_{0,1}^{-,+}(I_{0,1}^{-,+}+\hbar
\omega )^{2}+\Gamma ^{2}}\right\}.   
\end{align}
Combining Eqs. (10) , (12) , and (13), carrying out the sum over $n^{\prime }$,  and  
making the changes $n-1\rightarrow m\rightarrow n$ in one of the  sums, we obtain
\begin{equation} 
\sigma _{xy}^{nd}(\omega ) =\sigma _{0}\sum_{s,\eta ,n,\gamma ,\gamma
^{\prime }}
\left[ \frac{\left( f_{n+1}^{s\eta ,\gamma }-f_{n}^{s\eta ,\gamma
^{\prime }}\right) \,D_{n+1}^{s\eta ,\gamma }C_{n}^{s\eta ,\gamma ^{\prime }}%
}{I_{n+1,n}^{\gamma ,\gamma ^{\prime }}(I_{n+1,n}^{\gamma ,\gamma ^{\prime
}}+\hbar \omega -i\Gamma )}  - \frac{ 
(f_{n}^{s\eta ,\gamma }-f_{n+1}^{s\eta ,\gamma ^{\prime }}) 
\,C_{n}^{s\eta ,\gamma }D_{n+1}^{s\eta ,\gamma ^{\prime }}}{%
I_{n,n+1}^{\gamma ,\gamma ^{\prime }}(I_{n,n+1}^{\gamma ,\gamma ^{\prime
}}+\hbar \omega -i\Gamma )}  \right ]  
\end{equation} 
Now following the same procedure as the one adopted  for the nondiagonal
longitudinal conductivity (17), we obtain the imaginary part
of  the optical Hall conductivity as
\begin{align}
\notag
\Im \sigma _{xy}^{nd}(\omega )& =-\sigma _{0}\sum_{\eta ,s,n}\left\{ \frac{%
(f_{n}^{s\eta ,+}-f_{n+1}^{s\eta ,+}) \,\Gamma \,\left(
D_{n+1}^{s\eta ,+}C_{n}^{s\eta ,+}\right) ^{2}}{%
I_{n,n+1}^{+,+}\big[(I_{n,n+1}^{+,+}+\hbar \omega )^{2}+\Gamma ^{2}\big]}\right.
 -\frac{(f_{n}^{s\eta ,-}-f_{n+1}^{s\eta ,-}) \,\Gamma \,\left(
D_{n+1}^{s\eta ,-}C_{n}^{s\eta ,-}\right) ^{2}}{%
I_{n,n+1}^{-,-}\big[(-I_{n,n+1}^{-,-}+\hbar \omega )^{2}+\Gamma ^{2}\big]}  \\%
 &\hspace{0.99cm} -\frac{( f_{n+1}^{s\eta ,-}-f_{n}^{s\eta ,+})\,\Gamma \left(
D_{n+1}^{s\eta ,-}C_{n}^{s\eta ,+}\right) ^{2}}{%
I_{n+1,n}^{-,+}\big[(I_{n+1,n}^{-,+}+\hbar \omega )^{2}+\,\Gamma ^{2}\big]}   
 \left. +\frac{( f_{n}^{s\eta ,-}-f_{n+1}^{s\eta ,+}) \Gamma
\left( D_{n+1}^{s\eta ,+}C_{n}^{s\eta ,-}\right) ^{2}}{%
I_{n,n+1}^{-,+}\big[(I_{n,n+1}^{-,+}+\hbar \omega )^{2}+\Gamma ^{2}\big]}\right\} 
 \end{align}
 The results  for the $K^{\prime }$ valley can be  obtained by exchanging $C$ with $D$ in Eq. (19).
Following the same procedure as the one adopted for  $n\geq 1$, we obtain the imaginary part of the optical Hall
conductivity for the zeroth LL as
\begin{align}
\Im \sigma _{xy}^{nd}(\omega )& =-\sigma _{0}\sum_{\eta ,s}\left\{ \frac{%
\left[ f_{0}^{s\eta ,+}-f_{1}^{s\eta ,+}\right] \,\Gamma \,\left(
D_{1}^{s\eta ,+}\right) ^{2}}{I_{0,1}^{+,+}(I_{0,1}^{+,+}+\hbar \omega
)^{2}+\Gamma ^{2}}\right.   \label{18} 
 -\frac{\left[ f_{0}^{s\eta ,-}-f_{1}^{s\eta ,-}\right] \,\Gamma \,\left(
D_{1}^{s\eta ,-}\right) ^{2}}{I_{0,1}^{-,-}(-I_{0,1}^{-,-}+\hbar \omega
)^{2}+\Gamma ^{2}} \notag  \\ 
&\hspace*{1.5cm} -\frac{\left[ f_{1}^{s\eta ,-}-f_{0}^{s\eta ,+}\right] \,\Gamma \left(
D_{1}^{s\eta ,-}\right) ^{2}}{I_{1,0}^{-,+}(I_{1,0}^{-,+}+\hbar \omega
)^{2}+\,\Gamma ^{2}}   
 \left. +\frac{\left[ f_{0}^{s\eta ,-}-f_{1}^{s\eta ,+}\right] \Gamma
\left( D_{1}^{s\eta ,+}\right) ^{2}}{I_{0,1}^{-,+}(I_{0,1}^{-,+}+\hbar
\omega )^{2}+\Gamma ^{2}}\right\}   
\end{align}%
Similarly, by exchanging $D\rightarrow C$ in above Eq., we can be obtained
for the $K^{\prime }$-valley.

\section{Numerical results and discussion}

The energies of the positive branch levels in Eq. (2) are different than those
 of the negative branch due to different values of the SOC energy.
Because $\hslash \omega _{c}<<\Delta $, the {\it intraband} and {\it interband} 
transitions in WSe$_{2}$ belong to two completely different regimes:
the intraband transitions occur in the microwave-to-THz  
and the interband ones in the visible frequency range. We first consider the latter ones ($n^{\prime }=n\pm 1$).
Unlike graphene-like 2D systems, the huge band gap and strong SOC in  the WSe%
$_{2}$ spectrum have important implications for the peaks seen in $%
\Re \sigma _{xx}^{nd}(\omega )$ and $\Im \sigma _{xy}^{nd}(\omega )
$ as a function of the light frequency. This is shown in Fig. 5 for a temperature T = 5 K and 
a level broadening $\Gamma =0.2\sqrt{B}$ meV.
We take $B=30$ T in oder to have  well-resolved LLs.  
The black solid curve is for $E_{F}$ in the gap ($E_{F}=0.0$ eV), while  the red
dotted curve is for $E_{F}=0.892$ eV. This value falls between the $n=3$ and 
$n=4$ LLs. The optical selection rules allow $n$ to change by only 1. In
addition one needs to go from occupied to unoccupied states through the
absorption of  photons. For $E_{F}=0$ the peaks occur at $\hslash \omega
=E_{n+1}^{+}+E_{n}^{+}$ for integer $n$. The series of peaks corresponds to
allowed interband transitions in the LL structure. The peak spacing is
proportional to $B$ and can be seen  
even at weaker fields, say, for $B\geq 10$ Tesla. Similar to graphene-like 2D
systems, the spectral weight of the interband peaks is continuously
redistributed into the intraband peaks. This shows how the conductivity
changes as $E_F$  moves through the LLs.
In contrast to graphene in which the SOC is very weak, the strong spin splitting in WSe$_{2}$ leads  
to beating  patterns in $\sigma_{xx}$ as seen  in the right panel of Fig. 5.  
For low frequencies though, $\sigma_{xx}$ doesn't show any beating  pattern due to the well separated spin-up and spin-down states which do no mix  at these frequencies, cf. left panel of Fig. 5. 

\begin{figure}[t]
\includegraphics[width=0.4\textwidth,clip]{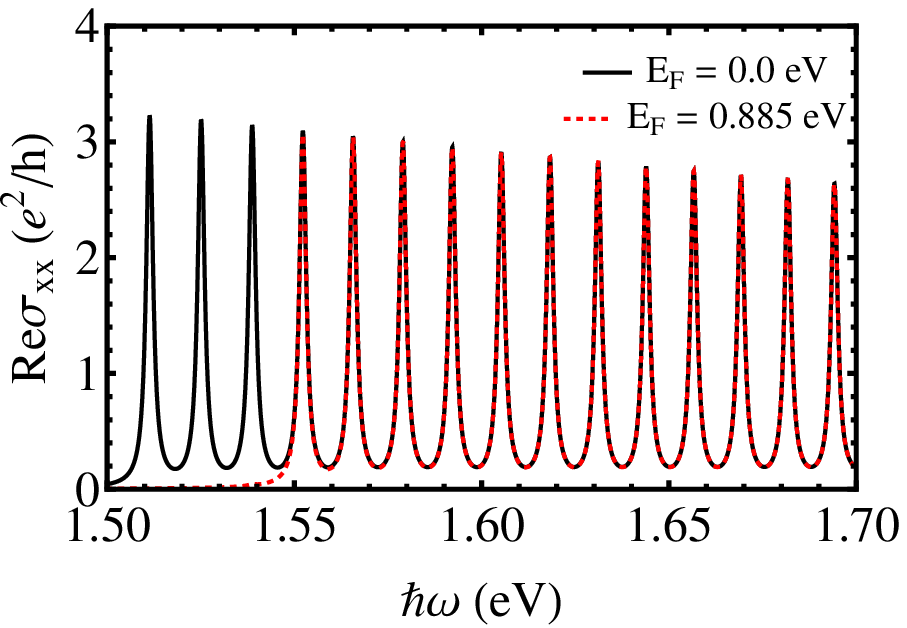}
\includegraphics[width=0.4\textwidth,clip]{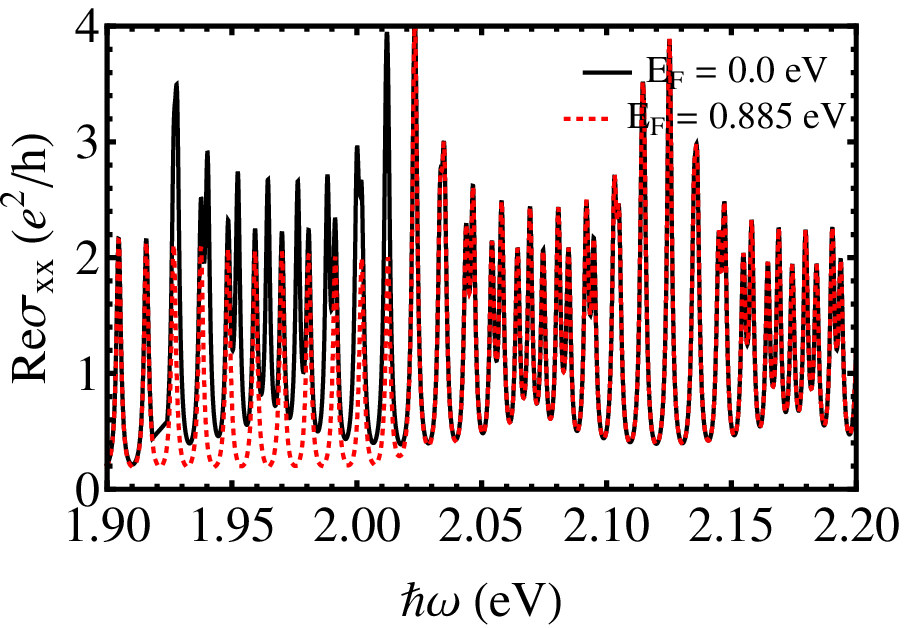}
\vspace*{-0.3cm}
\caption{(Coulour online) Real part of the longitudinal optical conductivity as a function
of the photon energy for a field $B=30$ Tesla. The two panels 
differ only in the  frequency range ($x$ axis).}  
\end{figure}

\begin{figure}[b]
\includegraphics[width=0.4\columnwidth,height=0.26\columnwidth]{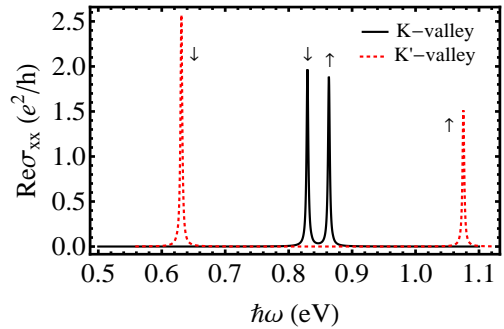}
\vspace*{-0.3cm}
\caption{(Coulour online) As in Fig. 5 but with  
only the $n=0$ LL taken into account. The spin assignment of the curves follows from Eq. (5).}  
\end{figure}

 In Fig. 6 we replot $\Re \sigma _{xx}^{nd}(\omega )$ with only the
$n=0$ LL taken into account. A magnetic control of the valley polarization 
can be clearly seen as the corresponding peaks in two different valleys 
appear at different frequencies. In addition to the valley-controlled transport, the spin splitting of the peaks into two in each valley is due to the strong SOC. The spin and valley splittings can be understood with the help of Eq. (5) and the corresponding energies. This is in line with 
the experimental realization of the valley-controlled dynamics of particles in WSe$_{2}$ at the Dirac point due to the Zeeman term \cite{GZ,AM}. In the pure Dirac case the spin and valley  peaks occur  at  the 
same frequency and hence cancel out perfectly in contrast to the four distinct peaks  in WSe$_{2}$  shown in Fig. 6. In  graphene only the first peak  occurs in this  conductivity component,
higher-order peaks are absent due to the cancellation just described. By contrast,  if we increase the value  $E_{F}$ so that it falls between the positive $n=3$ and $n=4$ LLs ($E_{F}=0.885$ eV) the
peaks don't cancel each other due to the asymmetric  SOC\ splittings in the two bands.
 We note that the lower peaks disappear as $E_{F}$ moves to higher LLs.

The peak structure just described above for $\Re \sigma _{xx}^{nd}(\omega )$
and $\Im \sigma _{xy}^{nd}(\omega )$ importantly affects their behavior for
right ($+$) and left ($-$) polarized light. For real experiments that probe
the (circular) polarization of resonant light, as in the case of the Kerr
and Faraday effects, one evaluates the quantity $\sigma _{\pm }(\omega )$
given by 
\begin{equation}
\sigma _{\pm }(\omega )=\Re \sigma _{xx}^{nd}(\omega )\mp \Im \sigma
_{xy}^{nd}(\omega ),  \label{20}
\end{equation}
with the upper (lower) sign corresponding to right (left) polarization \cite%
{VS,WA}.  In Fig. 7 we show $\sigma _{-}(\omega )$ (solid black curve)  
and $\sigma _{+}(\omega )$ (solid red curve) as functions of the frequency, both  for $E_{F}=0.0$ eV in the gap, using
the parameters of Fig. 5. As seen, there is a direct correspondence
between these results and those of Fig. 5. The heights of the peaks in $\sigma
_{+}(\omega )$ are much smaller than those  in $ \sigma _{-}(\omega )$. 
Similar to the behaviour of $\Re \sigma _{xx}^{nd}(\omega )$ and $\Re \sigma _{yx}^{nd}(\omega )$,
 spin  and valley splittings can be clearly seen  
in Fig. 8, in which the spin aspects of the curves is the same as in Fig. 6. We see four peaks due to the spin and valley splittings in accordance with Eq. (5) and in line 
with the obsrvation 
of valley-controlled dynamics of particles in WSe$_{2}$  \cite{GZ,AM}.  

The difference between $\sigma _{-}(\omega )$ and  $\sigma _{+}(\omega )$ is also reflected in the power absorption
spectrum given by 
\begin{equation}
\hspace*{-0.15cm}P(\omega )=(E/2)\left[ \sigma _{xx}(\omega )+\sigma
_{yy}(\omega )-i\sigma _{yx}(\omega )+i\sigma _{xy}(\omega )\right] .
\label{21}
\end{equation}
\begin{figure}[ht]
\vspace*{-0.2cm}
\includegraphics[width=0.6\columnwidth,height=0.35\columnwidth]{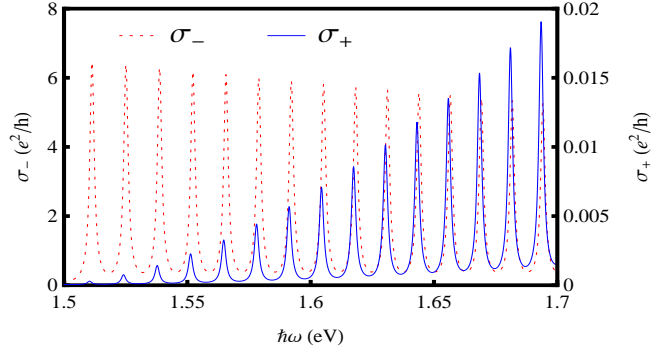}
\vspace*{-0.4cm}
\caption{(Coulour online) Real part of the right-polarized optical conductivity 
$\sigma _{+ }(\omega )$ and of the left-polarized one 
$\sigma _{- }(\omega )$  versus  photon energy for $E_F=0.0$ eV and  field $B=30$ Tesla.}
\end{figure}
\begin{figure}[ht]
\vspace*{-0.2cm}
\includegraphics[width=0.6\columnwidth,height=0.35\columnwidth]{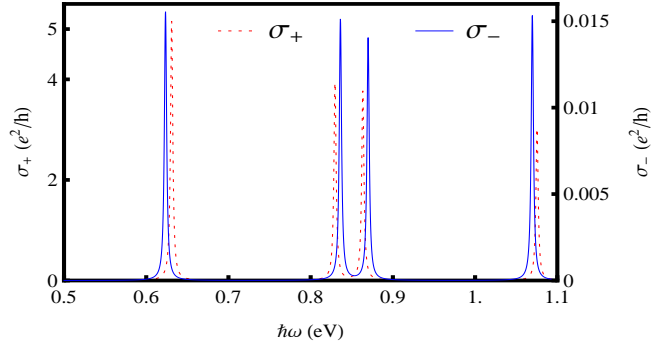}
\vspace*{-0.4cm}
\caption{As in Fig. 7 but  with only the $n=0$ LL taken into account.}
\end{figure}

\begin{figure}[ht]
\vspace*{-0.15cm}
\includegraphics[width=0.55\columnwidth,height=0.32\columnwidth]{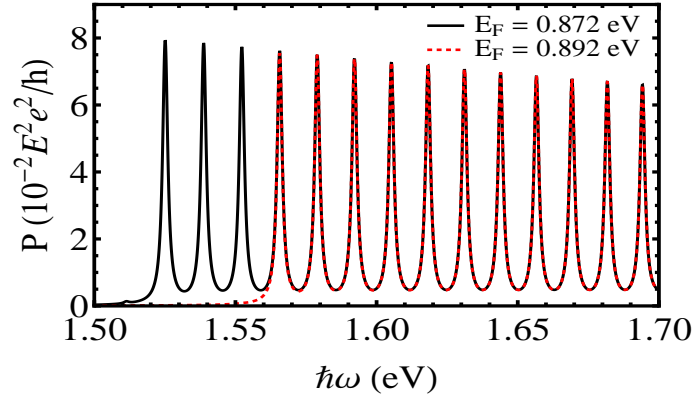}
\vspace*{-0.3cm}
\caption{(Coulour online) Power spectrum vs photon energy for an electric field $E=8$ V/nm,
 two values of $E_F$, and  field $B=30$ Tesla.}
\end{figure}
\begin{figure}[ht]
\vspace*{-0.35cm}
\includegraphics[width=0.5\columnwidth,height=0.32\columnwidth]{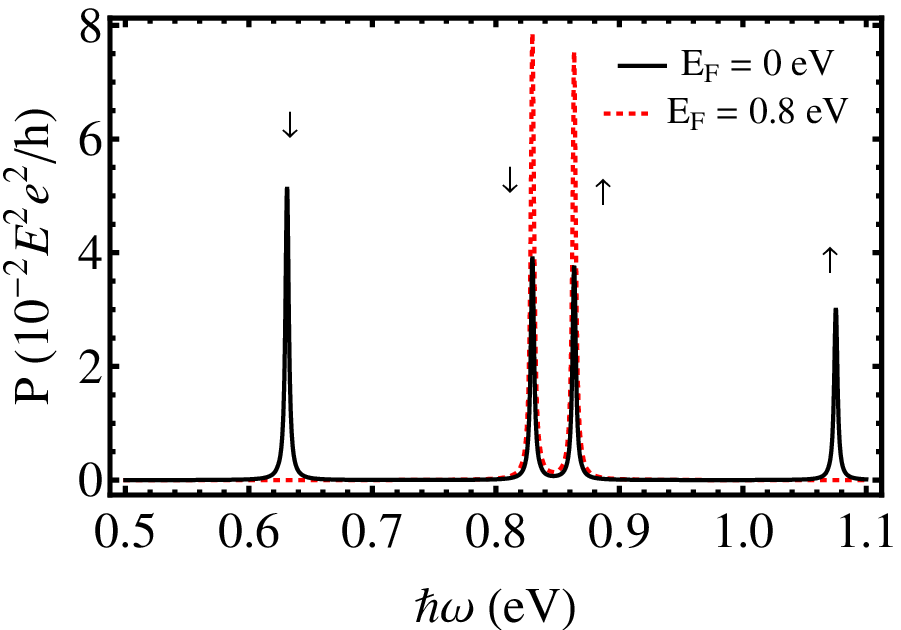}
\vspace*{-0.4cm}
\caption{(Coulour online) Power spectrum vs photon energy for an electric field $E=8$ V/nm,
two values of $E_F$, field $B=30$ Tesla, and 
only the $n=0$ LL taken into account. The spin assignment of the curves follows Eq. (5) and is identical to that of Fig. 6.
The outer peaks are for the $K^\prime$ valley and the inner ones for the $K$ valley}. 
\end{figure}
We remind that $\sigma _{\mu \nu }=\sigma _{\mu \nu }^{d}+\sigma _{\mu \nu
}^{nd}=\sigma _{\mu \nu }^{nd}$ since the component $\sigma _{\mu \mu
}^{d},\mu =x,y,$ vanishes. The component $\sigma _{yy}^{nd}(\omega )$ is
given by $\sigma _{xx}^{nd}(\omega )$ and $\Im \sigma _{xy}^{nd}(\omega
)=-\Im \sigma _{yx}^{nd}(\omega )$. The spectrum $P(\omega )$ is shown in
Fig. 9 as a function of the photon energy for two values of $E_{F}$. Given
that $\Im \sigma _{xy}^{nd}(\omega )$ is much smaller than $\Re \sigma
_{xx}^{nd}(\omega )$, cf. Fig. 5, the peaks in it are essentially the
same as those in the longitudinal optical conductivity. The absence of the $n\leq 4$
peaks for $E_{F}=0.982$ eV is due to Pauli blocking and consistent with Fig. 5.  
Similar to the $\Re \sigma _{xx}^{nd}(\omega )$ and $\Re \sigma _{yx}^{nd}(\omega )$,  spin  and valley splittings can be clearly seen  in Fig. 10, where we see four peaks due to these splittings  in accordance with Eq. (5). We find that by changing $E_F$  
from zero to a finite value, the power absorption peaks only for one valley,  as in Figs. 5-10.

Now we consider \textit{intraband} transitions between the $n$th and ($n+1$%
)th LLs {\it in the conduction band}, 
with $E_{F}>0$,  in which the energy change is much smaller than $%
E_{F}$. This involves large values of $n$ and is known as the semiclassical
limit of the magneto-optical conductivity in which $E_{F}$ is much larger
than $\hbar \omega _{c}$. Let us assume that $E_{F}\approx E_{n}^{+}$ lies
between the $n$th and ($n+1$)th LLs. The pertinent energy difference is $%
E_{n}^{+}-E_{n+1}^{+}=-\hslash \omega _{c}$. For such transitions we
obtain
\begin{equation}
\Re \sigma _{xx}^{nd}(\omega )=-\sigma _{0}\sum_{\eta ,s,n}\frac{(
f_{n}^{s\eta ,+}-f_{n+1}^{s\eta ,+}) \,\Gamma \,\left( D_{n+1}^{s\eta
,+}C_{n}^{s\eta ,+}\right) ^{2}}{I_{n,n+1}^{+,+}\big[(I_{n,n+1}^{+,+}+\hbar
\omega )^{2}+\Gamma ^{2}\big]}  \label{22}
\end{equation}

The real part of $\sigma _{xx}^{nd}(\omega )$ is shown in Fig. 11. 
As seen, the optical spectral weight under these curves
increases with $E_{F}$. These peaks lie in the range of microwave-to-THz
frequencies and their height is larger than that of the interband
transitions shown in Fig. 5-10. This is consistent with graphene or topological
insulators and other symmetric 2D systems in which the relevant spectral
weight increases with $E_{F}$, see, e.g., Fig. 7 of Ref. \cite{PJ}, and the
optical features appear in the THz regime only \cite{PJ,TY}. 
\begin{figure}[ht]
\vspace*{-0.2cm}
\includegraphics[width=0.45\columnwidth,height=0.3\columnwidth]{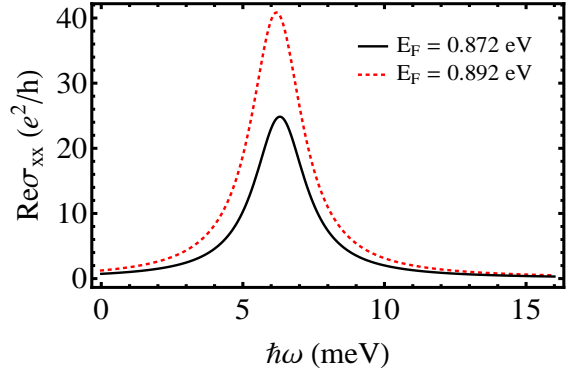}
\vspace*{-0.2cm}
\caption{(Coulour online) Intraband limit of the real part of the longitudinal optical
conductivity versus photon energy, two values of $E_{F}$, and 
$B=30$ Tesla. The energy 
$\hbar \omega$ is measured from 
the bottom of the conduction band.}
\end{figure}

\section{Summary}

\textit{\ } We studied  spin- and valley-controlled
magneto-optical transport properties of a WSe$_{2}$ monolayer  subject to a perpendicular magnetic field. 
We showed periodic oscillations with frequency of the conductivities due to the absorption of photons  
corresponding to LL transitions induced by the pertinent selection rules. Due to the large  direct band
gap of WSe$_{2}$ the conductivity peaks depend  linearly on $B$  and  reflect the equidistant LLs in each band. The {\it intraband} and {\it interband} optical transitions in WSe$_{2}$ belong to two completely different regimes:
the intraband one is in the microwave-to-THz range and the interband  one in the visible
frequency range.  The absorption peaks for the $n=0$ LL appear in between these two regimes and, as Figs. 6 and 10 demonstrate, a magnetic control of the valley and spin splittings is possible.
These  findings expand the horizon  of the electronic properties of 2D WSe$_{2}$
system and could be useful in the design of spintronic and
valleytronic optical devices.
\\

\vspace*{-0.2cm}
{\bf Acknowledgments}:
This work was supported by the the Canadian NSERC Grant No. OGP0121756.\\

\noindent Electronic addresses: $^{\star}$m.tahir06@alumni.imperial.ac.uk,
$^{\dag}$p.vasilopoulos@concordia.ca


\begin{thebibliography}{99} 

\bibitem{LY} L. Liao, Y.-C. Lin, M. Bao, R. Cheng, J. Bai, Y. Liu, Y. Qu, K.
L. Wang, Y. Huang, and X. Duan, Nature \textbf{467}, 305 (2010); F.
Schwierz, Nature Nanotechnology \textbf{5}, 487 (2010).

\bibitem{JT} J. Sone, T. Yamagami, Y. Aoki, K. Nakatsuji, and H. Hirayama,
New J. Phys. \textbf{16}, 095004 (2014); P. D. Padova, C. Ottaviani, C.
Quaresima, B. Olivieri, P. Imperatori, E. Salomon, T. Angot, L. Quagliano,
C. Romano, A. Vona, M. M.-Miranda, A. Generosi1, B. Paci, and G. L. Lay, 2D
Materials \textbf{1}, 021003 (2014).

\bibitem{ML} M. E. D\'{a}vila, L. Xian, S. Cahangirov, A. Rubio, and G. Le.
Lay, New J. Phys. \textbf{16}, 095002 (2014).

\bibitem{QK} Q. H. Wang, K. K.-Zadeh, A. Kis, J. N. Coleman, and M. S.
Strano, Nature Nanotechnology \textbf{7}, 699 (2012); A. K. Geim and I. V.
Grigorieva, Nature \textbf{499}, 419 (2013).

\bibitem{BA} B. Radisavljevic, A. Radenovic, J. Brivio, V. Giacometti and A.
Kis, Nature Nanotechnology \textbf{6}, 147 (2011); H. Fang, S. Chuang, T. C.
Chang, K. Takei, T. Takahashi, and A. Javey, Nano Lett. \textbf{12}, 3788
(2012); H. Wang, L. Yu, Y.-H. Lee, Y. Shi, A. Hsu, M. L. Chin, L.-J. Li, M.
Dubey, J. Kong, and T. Palacios, Nano Lett. \textbf{12}, 4674 (2012); M. S.
Fuhrer and J. Hone, Nature Nanotechnology \textbf{8}, 146 (2013).

\bibitem{DG} D. Xiao, G.-B. Liu, W. Feng, X. Xu, and W. Yao, Phys. Rev. Lett. 
\textbf{108}, 196802 (2012).

\bibitem{HW} H.-Z. Lu, W. Yao, D. Xiao, and S.-Q. Shen, Phys. Rev. Lett. 
\textbf{110}, 016806 (2013).

\bibitem{XF} X. Li, F. Zhang, and Q. Niu, Phys. Rev. Lett. \textbf{110},
066803 (2013).

\bibitem{AL} A. Splendiani, L. Sun, Y. Zhang, T. Li, J. Kim, C.-Y. Chim, G.
Galli, and F. Wang, Nano Lett. \textbf{10}, 1271 (2010).

\bibitem{JM} J. K. Ellis, M. J. Lucero, and G. E. Scuseria, Appl. Phys.
Lett. \textbf{99}, 261908 (2011).

\bibitem{HS} H. S. Lee, S.-W. Min, Y.-G. Chang, M. K. Park, T. Nam, H. Kim,
J. H. Kim, S. Ryu, and S. Im, Nano Lett. \textbf{12}, 3695 (2012).

\bibitem{ER} E. Cappelluti, R. Rold\'{a}n, J. A. Silva-Guill\'{e}n, P. Ordej%
\'{o}n, and F. Guinea, Phys. Rev. B \textbf{88}, 075409 (2013).

\bibitem{QKA} Q. H. Wang, K. Kalantar-Zadeh, A. Kis, J. N. Coleman, and M. S.
Strano, Nat. Nanotechnol. \textbf{7}, 699 (2012).

\bibitem{HA} H. Liu, A. T. Neal, and P. D. Ye, ACS Nano \textbf{6}, 8563
(2012).

\bibitem{YK} Y. Yoon, K. Ganapathi, and S. Salahudin, Nano Lett. \textbf{11}%
, 3768 (2011).

\bibitem{AH} A. M. Jones, H. Yu, J. S. Ross, P. Klement, N. J. Ghimire, J.
Yan, D. G. Mandrus, W. Yao, and X. Xu, Nature Physics \textbf{10}, 130
(2014).

\bibitem{GZ} G. Aivazian, Z. Gong, A. M. Jones, R.-L. Chu, J. Yan, D. G.
Mandrus, C. Zhang, D. Cobden, W. Yao, and X. Xu, Nature Phys. \textbf{11},
148 (2015).

\bibitem{AM} A. Srivastava, M. Sidler, A. V. Allain, D. S. Lembke, A. Kis,
and A. Imamo\u{g}lu, Nature Phys. \textbf{11}, 141 (2015).

\bibitem{HAS} H. C. P. Movva, A. Rai, S. Kang, K. Kim, B. Fallahazad, T.
Taniguchi, K. Watanabe, E. Tutuc, and S. K. Banerjee, ACS Nano, DOI:
10.1021/acsnano.5b04611 (2015).

\bibitem{AP} A. A. Mitioglu, P. Plochocka, \'{A}. G. del Aguila, P. C. M.
Christianen, G. Deligeorgis, S. Anghel, L. Kulyuk, and D. K. Maude, Nano
Lett. \textbf{15}, 4387 (2015).

\bibitem{VS} V. P. Gusynin, S. G. Sharapov, and J. P. Carbotte, Phys. Rev.
Lett. \textbf{98}, 157402 (2007); Z. Q. Li, E. A. Henriksen, Z. Jiang, Z.
Hao, M. C. Martin, P. Kim, H. L. Stormer, and D. N. Basov, Nat. Phys. 
\textbf{4}, 532 (2008); 
T. Stauber and N. M. R. Peres, J. Phys.: Condens. Matter 
\textbf{20}, 055002 (2008). 

\bibitem{WA} W.-K. Tse and A. H. MacDonald, Phys. Rev. B \textbf{84}, 205327
(2011); I. Garate and M. Franz, Phys. Rev. B \textbf{84}, 045403 (2011); D.
K. Efimkin and Y. E. Lozovik, Phys. Rev. B \textbf{87}, 245416 (2013); A
Ullah and K Sabeeh, J. Phys.: Condens. Matter \textbf{26}, 505303 (2014); M.
Lasia and L. Brey, Phys. Rev. B \textbf{90}, 075417 (2014).

\bibitem{ZJ} Z. Li and J. P. Carbotte, Phys. Rev. B \textbf{86}, 205425
(2012). 

\bibitem{LC} L. Stille, C. J. Tabert, and E. J. Nicol, Phys. Rev. B \textbf{%
86}, 195405 (2012).

%

\bibitem{MP} M. Tahir, P. Vasilopoulos, and F. M. Peeters, Phys. Rev. B 
\textbf{92}, 045420 (2015).

\bibitem{YT} Y. Zheng and T. Ando, Phys. Rev. B \textbf{65}, 245420 (2004).

\bibitem{Wang}  X. F. Wang and  P. Vasilopoulos, Phys. Rev. B{\bf 72}, 085344 (2005); {\it ibid.} 085313 (2003).

\bibitem{MK} M. Charbonneau, K. M. Van Vliet, and P. Vasilopoulos, J. Math.
Phys. \textbf{23}, 318 (1982).

\bibitem{PJ} P. E. C. Ashby and J. P. Carbotte, Phys. Rev. B \textbf{87},
245131 (2013). 

\bibitem{TY} T. Morimoto, Y. Hatsugai, and H. Aoki, Phys. Rev. Lett. \textbf{%
103}, 116803 (2009).
\end{thebibliography}
\end{document}